\documentclass[%
aip,
amsmath,amssymb,
reprint,%
]{revtex4-2}
\usepackage[
pdffitwindow=true,
colorlinks=true,
frenchlinks=false,
linkcolor=blue,
anchorcolor=blue,
citecolor=blue,
filecolor=blue,
urlcolor=blue,
bookmarks=true,
bookmarksopen=true,
bookmarksnumbered=true,
bookmarksopenlevel=1,
plainpages=false,
pdfpagelayout=TwoPageLeft,
pdfpagelabels=true,
breaklinks
]{hyperref}

\usepackage[per-mode=symbol,separate-uncertainty]{siunitx}
\usepackage{graphicx}
\usepackage{dcolumn}
\usepackage{url}
\usepackage{color}
\usepackage{graphicx,wrapfig,lipsum}
\usepackage{bm}
\usepackage[export]{adjustbox}
\usepackage[english]{babel}

\usepackage{color}
\usepackage{ulem}

\usepackage{lipsum}
\usepackage{amsmath}
\bibpunct{[}{]}{,}{n}{}{}
\usepackage{chemformula}

\begin{document}
	\title{Temperature dependence of the magnon-phonon coupling in yttrium iron garnet/gadolinium gallium garnet high overtone bulk acoustic resonators 
	}
	
		\author{J.~Weber}
	\email{johannes.weber@wmi.badw.de}
	\thanks{Contributed equally to this work}
	\affiliation{Walther-Mei{\ss}ner-Institut, Bayerische Akademie der Wissenschaften, 85748 Garching, Germany}
	\affiliation{Technical University of Munich, TUM School of Natural Sciences, Physics Department, 85748 Garching, Germany}

	\author{M.~M\"uller}
	\thanks{Contributed equally to this work}
	\affiliation{Walther-Mei{\ss}ner-Institut, Bayerische Akademie der Wissenschaften, 85748 Garching, Germany}
	\affiliation{Technical University of Munich, TUM School of Natural Sciences, Physics Department, 85748 Garching, Germany}
        \author{M.~Cherkasskii}
	\affiliation{Institute for Theoretical Solid State Physics, RWTH Aachen University, 52074 Aachen, Germany}
	
    \author{S.~Geprägs}
	\affiliation{Walther-Mei{\ss}ner-Institut, Bayerische Akademie der Wissenschaften, 85748 Garching, Germany}
	
	\author{R.~Gross}
	\affiliation{Walther-Mei{\ss}ner-Institut, Bayerische Akademie der Wissenschaften, 85748 Garching, Germany}
	\affiliation{Technical University of Munich, TUM School of Natural Sciences, Physics Department, 85748 Garching, Germany}
	\affiliation{Munich Center for Quantum Science and Technology (MCQST), 80799 Munich, Germany}

	\author{C.H.~Back}
        \affiliation{Technical University of Munich, TUM School of Natural Sciences, Physics Department, 85748 Garching, Germany}
        \affiliation{Munich Center for Quantum Science and Technology (MCQST), 80799 Munich, Germany}
	\affiliation{Centre for Quantum Engineering (ZQE), Technical University of Munich, 85748 Garching, Germany}

	\author{S.T.B.~Goennenwein}
	\affiliation{Department of Physics, University of Konstanz, 78457 Konstanz, Germany}
	\author{S.~Viola Kusminskiy}
	\affiliation{Institute for Theoretical Solid State Physics, RWTH Aachen University, 52074 Aachen, Germany}
	\affiliation{Max Planck Institute for the Science of Light, 91058 Erlangen, Germany}

	\author{M.~Althammer}
	
	\affiliation{Walther-Mei{\ss}ner-Institut, Bayerische Akademie der Wissenschaften, 85748 Garching, Germany}
	\affiliation{Technical University of Munich, TUM School of Natural Sciences, Physics Department, 85748 Garching, Germany}
	\author{H.~Huebl}
	\email{huebl@wmi.badw.de}
	\affiliation{Walther-Mei{\ss}ner-Institut, Bayerische Akademie der Wissenschaften, 85748 Garching, Germany}
	\affiliation{Technical University of Munich, TUM School of Natural Sciences, Physics Department, 85748 Garching, Germany}
	\affiliation{Munich Center for Quantum Science and Technology (MCQST), 80799 Munich, Germany}
\begin{abstract}
We experimentally study the temperature dependence of the magnon-phonon coupling in a yttrium iron garnet (YIG)/gadolinium gallium garnet (GGG) heterostructure. More specifically,  we use broadband ferromagnetic resonance to investigate the magneto-elastic coupling between the Kittel mode of a YIG thin film and the transverse acoustic phonon modes of the YIG/GGG high overtone bulk acoustic wave acoustic resonator for in and out-of-plane field directions in the temperature range between  $T=\SI{5}{K}$ and $\SI{300}{K}$. We find that for a magnetic field applied normal to the film surface, magneto-elastic coupling decreases with decreasing temperature, whereas it increases for the in-plane magnetic field configuration. The observed temperature dependence differs from earlier observations on bulk YIG samples, which might be due to the temperature dependent stress imposed by the GGG substrate. 
\end{abstract}
\maketitle

\section{Introduction}	 
The coupling between spin and lattice degrees of freedom is of fundamental importance to magnetic materials. It allows to excite magnetization dynamics \cite{Weiler2011, Weiler2012, Gowtham2016, Ku2020, Kuß2021, Kuß2021a, Kikkawa2016, Hashimoto2018a, Zhang2020a, Hatanaka2022, Künstle2025}, affects the damping properties of the magnetization dynamics \cite{Spencer1958, Widom2010, Vittoria2010, Rossi2005, Strongin1976, Kobayashi1973}, enables the control of the magnetization direction \cite{Weiler2009, Uchida2011, Weiler2011, Weiler2012, Thevenard2014,Thevenard2013, Geprägs2010, Geprägs2014, Brandlmeier2008}, and allows to create perpendicular magnetic anisotropy \cite{Weiler2011, Dreher2012,Qurat-Ul-Ain2020,Bi2010,An2016a}. The spin-lattice coupling concomitantly results in a finite interaction of the associated elementary excitations, i.e., magnons and phonons, giving rise to the formation of hybridized modes \cite{Yahiro2020, Sukhanov2019, Hayashi2018, Brataas2020, Kikkawa2016}. Recently, the resonant interaction of magnons and phonons mediated by magneto-elastic coupling regained interest in the context of understanding magnetization damping, excitation of helical phonons, strong coupling phenomena, as well as applications in quantum sensing and transduction  \cite{Streib2018,Sato2021, An2020, An2022,Schlitz2022, Sharma2022a, Bittencourt2022, Wachter2021, Potts2021, An2023,Keshtgar2014, Flebus2017, Sharma2022,Graf2021,Engelhardt2022a,Hatanaka2022, Künstle2025, Kim2025,Shen2025,Cui2023,Matsumoto2024}. In this context, many experiments explore bulk acoustic wave (BAW) resonators based on gadolinium gallium garnet (\ch{Gd3Ga5O12}, GGG) substrates combined with a layer of ferrimagnetic yttrium iron garnet (\ch{Y3Fe5O12}, YIG) \cite{An2020, An2022, Schlitz2022}, mostly due to the exceptional damping properties of YIG \cite{Maier-Flaig2017b,Jermain2017, Spencer1959, Klingler2017}. However, the majority of experiments so far have focused on room-temperature properties, because the magnetization damping of YIG increases at cryogenic temperatures \cite{Maier-Flaig2017b,Jermain2017, Lachance-Quirion2019a,Serha2025}. At the same time, the acoustic losses  \cite{Transducers1964, Landau1937,Selfcife2022, müller2023} are strongly reduced at low temperatures what could result in overall improved cooperativities $C$. The magneto-elastic coupling in YIG is also temperature dependent, and, moreover, anisotropic, since the interaction reflects the symmetry of the lattice\,\cite{Cullen1963, Hansen, Barangi2015}.
Hence, the precise temperature dependence of the magnetoelastic interaction in garnet-based heterostructures is of great interest for magneto-acoustic hybrids for low-temperature and potential quantum applications \cite{Graf2021,Engelhardt2022a}. 

Here, we explore the coupling between the magnetic excitations of YIG and the elastic excitations of bulk acoustic resonator modes of the YIG/GGG heterostructure system between $\SI{5}{K}$ and room temperature. Using broadband ferromagnetic resonance spectroscopy, we extract the magneto-elastic coupling strength and anisotropy, as well as the magnetic and acoustic damping rates. We find that the damping of the magnetization dynamics due to magneto-elastic coupling is strongly temperature-dependent and anisotropic.

\section{Experimental Details}

\begin{figure}[h]
    \centering
    \includegraphics[width=0.43\textwidth]{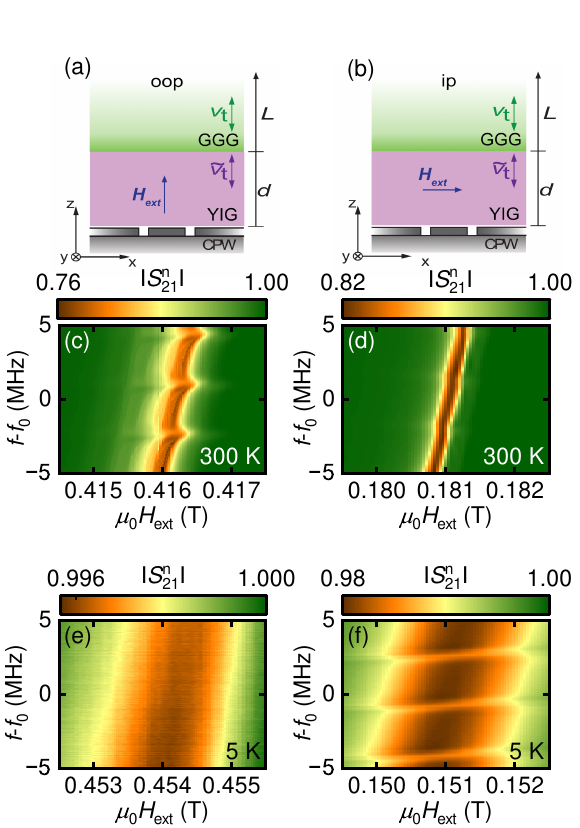}
    \caption{(a),(b) Coordinate system as well as schematic of the sample composed of a ferrimagnetic YIG thin film with a thickness $d$ on a GGG substrate with thickness $L$, mounted on a coplanar waveguide (CPW). The external magnetic field $H_\mathrm{ext}$ is either applied out of the thin film plane (oop) along the $z$- direction (panels (a), (c), (e)) or in-plane (ip) along the $x$-direction (panels (b), (d), (f)). The two different sound velocities are referred to as $\Tilde{v}_\mathrm{t}$ (YIG) and $v_\mathrm{t}$ (GGG). 
    Panels (c)-(f) show the normalized complex transmission amplitude $\lvert S_{21}^\mathrm{n} \rvert$ as a function of the external magnetic field $H_\mathrm{ext}$ and the microwave frequency $f-f_0$ with $f_0 = \SI{7}{GHz}$ for $T = \SI{300}{K}$ and $\SI{5}{K}$ as well as for the ip and oop magnetic field configuration.
    }
    \label{fig:colormaps}
\end{figure}

In our experiments, we studied a BAW resonator comprised of a $d = \SI{220}{nm}$ thick, (111) YIG film grown via liquid phase epitaxy (LPE) on a crystalline GGG substrate with a thickness of $L = \SI{530}{\micro\meter}$. As sketched in Fig.\,\ref{fig:colormaps}, the sample is mounted face-down onto a coplanar waveguide (CPW), which is connected to a vector network analyzer (see Ref.\,\cite{Selfcife2022,müller2023} for details). We perform temperature-dependent broadband ferromagnetic resonance experiments (bbFMR), i.e., we measure the complex transmission parameter $S_{21}$ as a function of frequency and applied magnetic field $H_\mathrm{ext}$ for various sample temperatures ranging from $\SI{5}{K}$ to room temperature \cite{Maier-Flaig2018}. To this end, the sample with the CPW is mounted in the variable-temperature insert of a liquid helium cryostat. The external magnetic field is applied either parallel to the surface normal and perpendicular to the transmission line of the CPW (out-of-plane, oop) or perpendicular to both the surface normal and the transmission line of the CPW (in-plane, ip) (see Fig.\,\ref{fig:colormaps} (a) and (b)).

Figure\,\ref{fig:colormaps} (c)-(f) displays the broad-band ferromagnetic resonance data in form of the normalized absolute transmission amplitude $\lvert S_{21}^\mathrm{n} \rvert = \lvert S_{21}/S_{21}^0\rvert$, with $S_{21}^{0}$ as a cut at constant field off-resonant of the Kittel mode, as a function of the microwave frequency $f-f_0$ with $f_0= \SI{7}{GHz}$ and the applied magnetic field $\mu_0 \lvert \mathbf{H_\mathrm{ext}}\rvert$ for $\mathbf{H_\mathrm{ext}}$ along the $z$-direction (oop, see panel (c), (e)) and $x$-direction (ip, see panel (d), (f)) for $T=\SI{5}{K}$ and $\SI{300}{K}$, respectively \footnote{We experimentally determine $S_{21}^0$ by measuring the complex transmission parameter $S_{21}$ far off-resonant with the FMR mode.}.  This high-resolution zoom into the broadband microwave absorption data shows the Kittel mode as an absorption signature in brown color. The linear relation between the resonance frequency and the applied magnetic field originates from the gyromagnetic ratio of YIG. The magnetic field range of panel (c)-(f) is $\SI{3}{mT}$, however, the center field varies due to the impact of the shape anisotropy and the change of the saturation magnetization with temperature \cite{Dionne2014}. As evident from Fig.\;\ref{fig:colormaps}, we find additional absorption signatures which are periodic in frequency, indicating the coupling of the Kittel mode to the resonant elastic excitations of the bulk acoustic wave resonator. Their frequency separation or free spectral range (FSR) of $f_\mathrm{FSR}=\SI{3.57}{MHz}$ at $\SI{5}{K}$ is related to the shear wave velocities of the elastic excitations and thickness of the total layer stack via $f_\mathrm{FSR}=1/(2 (d/\Tilde{v_\mathrm{t}} + L/v_\mathrm{t}))$ \cite{Sato2021}. 
Here, $d = \SI{220}{nm}$ and $\Tilde{v}_\mathrm{t} = \SI{3843}{ms^{-1}}$ \cite{Sato2021} are the thickness and the transverse acoustic sound velocity of the YIG thin film, and $L = \SI{530}{\micro\meter}$ and $v_\mathrm{t} = \SI{3568}{ms^{-1}}$ \cite{Sato2021} are the thickness and transverse acoustic sound velocity of the GGG substrate. As $L\gg d$,  the acoustic properties of the layer stack are dominated by the elastic properties of GGG. Therefore, we find good agreement of the measured FSR with the calculated one using the elastic properties of GGG. By analyzing the FMR linewidth at a frequency off-resonant with the acoustic modes, we determine the change of linewidth with temperature. The FMR linewidth clearly increases at low temperatures in agreement with literature \cite{Sato2021, Maier-Flaig2017b, Zhang2014,Spencer1959} (see also Fig.\,\ref{fig:eta-kappa-B}).

In addition, we find a remarkable difference in the visibility of the frequency-periodic features due to magneto-elastic coupling for experiments with in-plane and out-of-plane field direction. At $T=\SI{300}{K}$, 

we observe a clear signature of magnon-phonon coupling for the out-of-plane magnetic field configuration in agreement with the room-temperature results in Ref.\,\cite{An2023}, while such a signature is absent at $T=\SI{5}{K}$. The opposite is the case for the in-plane magnetic field configuration. Since the visibility of this signature depends on the magneto-elastic coupling strength, the data in Fig.\,\ref{fig:colormaps} suggest that the anisotropy of the magneto-elastic interaction and the magnitude of its components change as a function of temperature. 

\section{Modeling of magnon-phonon coupling and comparison with data}

In order to understand the interaction between the magnetic excitations confined to the thin YIG film and the elastic excitations extending over the entire sample stack comprised of the YIG film and the GGG substrate, several factors have to be considered. In particular, one has to take into account the acoustic mode matching, the elastic coupling at the interface, the mode matching between the relevant acoustic and magnetic modes, the details of the sample geometry, as well as the material parameters describing the magnetization dynamics in the thin film, the acoustic excitations in the layer stack, and the magneto-elastic coupling giving rise to the magnon-phonon coupling \cite{Streib2018,Sato2021,Cherkasskii2025, Rückriegel2014,Gurevich2020}.
For the quantitative analysis of the relevant parameters, we employ the models developed in\,\cite{Sato2021, Cherkasskii2025}. In our experimental situation, the normal on the YIG/GGG bilayer is oriented approximately along the [111] direction of YIG. Furthermore, we need to account for deviations in the phonon propagation with respect to the crystallographic [111]-direction, originating e.g. from a finite miscut of the GGG substrate \cite{Selfcife2022}. We fit our data to the description presented in Ref.\,\cite{Cherkasskii2025}, which accounts for those details. In Appendix \ref{Appendix: fitting function}, we extend this approach to account for in-plane magnetic fields. Thereby, we relate the Polder susceptibility to the experimentally measured data using \cite{Schoen2015}
\begin{equation}
\label{eq:FitTransmission}
    S_{21}(f, H_\mathrm{ext}) =A + D H_\mathrm{ext} + \sum_{n}^N E e^{i\phi} \chi_n(f, H_\mathrm{ext}).
\end{equation}
Here, $A$ and $D$ are parameters that account for a complex background of the microwave transmission up to linear order in the applied external magnetic field $H_\mathrm{ext}$. In addition, $E$ is the coupling between the sample and the magnetic circuit \footnote{Since we are evaluating the data in a small frequency window, we assume that  $E$ is constant.}  and $\phi$ corrects phase shifts, which originate from the experimental setup.
Equation\,\eqref{eq:FitTransmission} allows us to quantitatively analyze our data. In particular, we perform a least mean square 2D-fit of the complex transmission data recorded as function of the frequency and the applied magnetic field for all temperatures, which allows us to determine the parameters associated with the microwave detection setup ($A$, $D$, $E$ and $\phi$) as well as the Polder susceptibility providing insights into the magnetization dynamics, the coupling to the elastic excitations of the acoustic resonator, as well as the acoustic damping.

\subsection{Comparison of model predictions and data}

\begin{figure}[tb]
    \centering
    \includegraphics[width=0.5\textwidth]{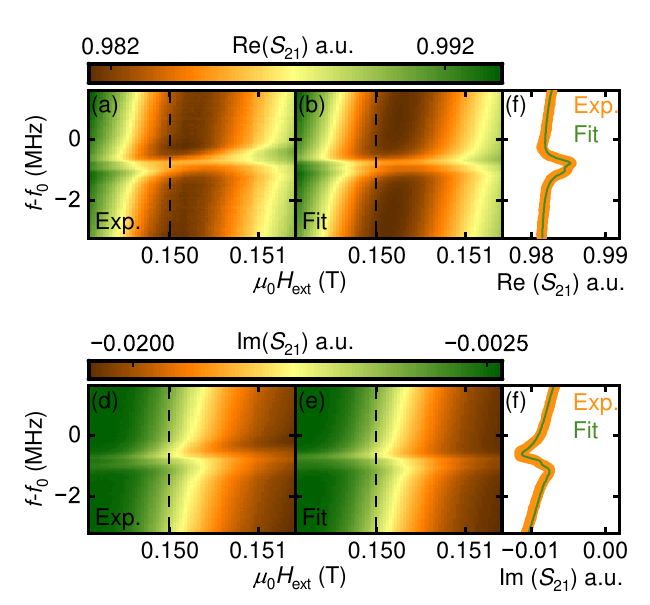}
    \caption{Panels (a) and (d) show the real and imaginary part of the transmission parameter $S_{21}$ of a FMR measurement at 5\,K with the magnetic field applied within the plane of the YIG thin film. The avoided crossing with the transverse acoustic phonon mode is clearly visible as a horizontal line in the middle of the panels. Panels (b) and (e) display the result of a simultaneous 2D fit of the real and imaginary parts. Panels (c) and (f) show two exemplary line cuts at $\SI{0.1496}{T}$ represented by the black dashed vertical lines in the color plots. The orange line is the measured data, and the green line is the fit data.}
    \label{fig:5K-fitdata}
\end{figure}

Figure \ref{fig:5K-fitdata} compares the experimental data with the fit result. This asserts the adequacy of the fit model, which reproduces the experimental data with high accuracy. 
The susceptibility derived in Refs.\,\cite{Sato2021, Cherkasskii2025} and explicitly stated for our analysis in App.\,\ref{Appendix: fitting function} is used for the fitting and depends on multiple parameters. Practically, we optimize the effective magnetization $M_\mathrm{eff}^\mathrm{ip,oop}$, the magnetic damping rate $\kappa_\mathrm{s}$, the acoustic damping rate $\eta_\mathrm{a}$ and the magneto-elastic coupling parameter $B_\mathrm{ip/ oop}$. The following parameters are fixed:  (i) the transverse phonon propagation velocities $v_\mathrm{t}^\mathrm{x,y}= \SI{3568}{{m}/{s}}$ which are degenerate for a propagation along the [111]-direction in GGG \footnote{In our experiment, the surface normal on the YIG film is not perfectly aligned with the [111]-direction of the YIG and GGG crystal due to the presence of a small, but finite miscut of \SI{0.003}{\degree}. Ref.\,\cite{Cherkasskii2025} as well as the adaption for the magnetic field aligned in-plane accounts for the resulting lifting in the degeneracy of the propagation velocities by distinguishing two shear wave velocities $v_\mathrm{t, x}$ and $v_\mathrm{t, y}$ with a velocity difference  $v_\mathrm{t, x}-v_\mathrm{t, y} = \Delta v$}, (ii)  the mass density $\rho = \SI{7070}{{kg}/{m^3}}$\,\cite{Sato2021}, (iii) the shear modulus $C_{44} = \SI{90.8}{GPa}$\,\cite{Kleszczewski1988}, and (iv) the thickness $L$ of the GGG substrate. For the YIG layer, we fix (i) the transverse acoustic sound velocity $\Tilde{v_\mathrm{t} = \SI{3843}{m/s}}$ (ii) the mass density $\Tilde{\rho} = \SI{5170}{{kg}/{m^3}}$, (iii) the shear modulus $\Tilde{C}_{44} = \SI{76.35}{GPa}$\,\cite{Gurevich2020} and (iv) the thickness $d$ of the YIG thin film. In addition, we assume that the acoustic damping is identical for GGG and YIG $\eta_\mathrm{a}=\Tilde{\eta}_\mathrm{a}$. Last but not least, the susceptibility also includes the gyromagnetic ratio, which is determined independently and set to ${\gamma}/{2\pi} = \SI{28}{{GHz}/{T}}$. 

The exact elastic resonance frequencies of the bulk acoustic resonator are temperature dependent due to the thermal expansion of the layer stack and subtle changes in the elastic parameters of the material \cite{müller2023}. We account for this in our fit by introducing an artificial frequency offset $\Delta f$ that corrects the resonance frequency of the transverse acoustic phonon modes to match the frequencies observed in experiment. 

Additionally, we have four (complex) fit parameters that account for $A$, $D$, $E$, and $\phi$ in Eq.\,\eqref{eq:FitTransmission}, which are approximately temperature independent.

Figure\;\ref{fig:5K-fitdata} (a) and (d) present the real and imaginary part of the transmission parameter $S_{21}$ of a bbFMR measurement at $\SI{5}{K}$ in the ip magnetic field configuration. The data are zoomed in on a single avoided crossing between the two transverse acoustic phonon modes and the Kittel mode. Panels (b) and (e) show the corresponding fit of Eq.\,\eqref{eq:FitTransmission} to the experimental data, while panels (c) and (f) display an exemplary cut at $\mu_0H_\mathrm{ext} = \SI{0.1496}{T}$, (see black dashed lines in panels (a), (b), (d) and (e)). The measurement data, represented by the orange lines in panels (c) and (f), agree well with the fit presented as a green line. Furthermore, the real and imaginary parts of the transmission parameter $S_{21}$ agree quantitatively with our theoretical model. In addition to the data presented in Fig.\,\ref{fig:colormaps}, we analyze further experimental data recorded for temperatures between $T=\SI{5}{K}$ and $\SI{300}{K}$  using this procedure for magnetic fields aligned along the normal and in-plane direction of the sample.  The extracted parameters are discussed in the following.

\subsection{Temperature dependence of material parameters}

\begin{figure}[tbh]
    \centering
    \includegraphics[width=0.5\textwidth]{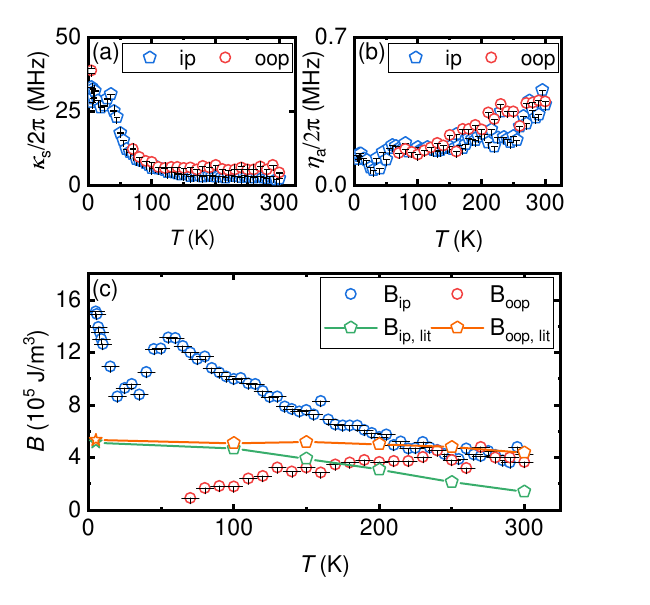}
    \caption{{Magnetic $\kappa_\mathrm{s}$ (a) and acoustic $\eta_\mathrm{a}$ (b) damping rates of the YIG/GGG high overtone bulk acoustic wave resonator as a function of temperature for both the ip and oop magnetic field configuration.
    Panel (c) shows the temperature dependence of the coupling parameters $B_\mathrm{ip}$ and $B_\mathrm{oop}$ (blue and red circles, respectively) obtained from fitting the experimental data. Also shown are $B_\mathrm{ip, lit}$ and $B_\mathrm{oop, lit}$ derived from literature data on bulk YIG. The two stars mark the extrapolated values at $\SI{5}{K}$.} The error bars originate from the covariance matrix of the least mean square fit.}
    \label{fig:eta-kappa-B}
\end{figure}

We next discuss the temperature dependence of the material parameters obtained from fitting the experimental data. Figure\;\ref{fig:eta-kappa-B} (a) and (b) show the temperature dependence of the magnetic and acoustic damping rates, $\kappa_\mathrm{s}$ and $\eta_a$, respectively,  for the ip (blue open circles) and oop  (red open circles) magnetic field configuration. Here, the magnetic damping rate does not include the damping originating from the coupling to the elastic excitations of the BAW resonator. The magnetic damping $\kappa_\mathrm{s}$ continuously increases towards lower temperatures. Note that we measured and evaluated the data for the oop magnetic field configuration down to $\SI{70}{K}$, where a coupling between the magnetic and elastic subsystems still can be resolved in our experiments. The damping rate for the oop configuration at $\SI{5}{K}$ is extracted by fitting a  Polder susceptibility without coupling to the elastic excitations \cite{Polder1949}.
For the ip magnetic field configuration, we find a monotonic increase from $T = \SI{300}{K}$ to $\SI{30}{K}$ followed by a plateau for even lower temperatures. This observation is commonly reported for YIG and attributed to the coupling of the magnetization dynamics to impurities \cite{Spencer1959,Maier-Flaig2017b, woltersdorf09}. 
Panel (b) shows the expected monotonic increase of the acoustic damping rate $\eta_\mathrm{a}$ measured at $f_0 = \SI{7}{GHz}$ with increasing temperature. This can be attributed to the scattering of the excited transverse acoustic phonons with thermal phonons\,\cite{Wolfe1998,Landau1937}. 
Our data suggests that both the acoustic and magnetic relaxation behaviors are independent of the measurement geometry (ip or oop), which can be understood in the sense that they represent bulk properties of the magnetic and acoustic subsystem.

In addition to the relaxation rates, we extract the magneto-elastic coupling constants $B_\mathrm{oop}$ and $B_\mathrm{ip}$ by fitting the data (see Fig.\,\ref{fig:eta-kappa-B} (c)). These constants are linked to the magneto-elastic coupling constants of the crystalline systems via $B_\mathrm{oop}={(2B_1 + B_2)}/{3}$ and $B_\mathrm{ip} ={\sqrt{2}\left(B_2 - B_1\right)}/{3}$, respectively \cite{Iida1964}. The latter are related to the magnetostriction constants of a cubic system via $\lambda_{100} = -\frac{2}{3}\frac{B_1}{C_{11} - C_{12}}$ and $\lambda_{111} = -\frac{B_2}{3C_{44}}$ (cf. Ref.\,\cite{Fritsch2012}). Figure\;\ref{fig:eta-kappa-B} (c) also includes data for bulk YIG from Cullen \textit{et al.}\,\cite{Cullen1963, Hansen} obtained from the published temperature-dependent magnetostriction parameters $\lambda_{100}$ and $\lambda_{111}$ for the temperature range between $\SI{100}{K}$ and $\SI{300}{K}$.  For the oop magnetic field configuration, we find reasonable quantitative agreement within 25\% with the literature values for $T\geq\SI{150}{K}$. Below $T\approx\SI{150}{K}$, we observe a reduction in $B_\mathrm{oop}$ down to a level, where our fit is compatible with $B_\mathrm{oop}=0$. For the ip field orientation, we determine a similar temperature dependence compared to Refs.\,\cite{Cullen1963, Hansen}, however, our magneto-elastic parameter is  approximately a factor of $2$ larger. Moreover, between $\SI{5}{K}$ and $\SI{100}{K}$, we observe an additional increase of $B_\mathrm{ip}$ up to $\SI{13.12e5}{\frac{J}{m^3}}$. Note that the data by Cullen \textit{et al.}\,\cite{Cullen1963} extrapolates the magneto-elastic constant in this temperature regime using the theory of magnetostriction in a cubic insulator applied to a ferrimagnet and using the sublattice magnetization data of C. Robert\,\cite{Robert1961}. Interestingly, while our data is in rough agreement with the observations for bulk YIG for elevated temperatures, we find discrepancies for lower $T$. We attribute this to the different growth techniques employed. Our $\SI{220}{nm}$ thick YIG film is grown using liquid-phase epitaxy techniques, which involve solvents such as lead and boron oxides, potentially leading to a residual contamination of the magnetic film. In addition, we expect a difference in the expansion coefficient between YIG and GGG, resulting in additional strain at lower temperatures. Together, these aspects could explain the difference in the observed magneto-elastic coupling coefficient. 

\begin{figure}[tbh]
    \centering
    \includegraphics[width=0.5\textwidth]{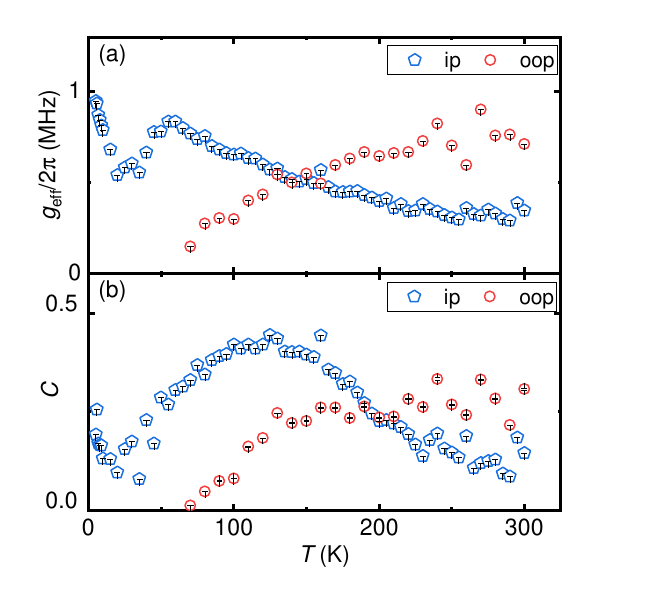}
    \caption{(a) Effective coupling strength $g_\mathrm{eff}$ as a function of temperature calculated from Eqs.\,\eqref{eq:olivierkleinip},\,\eqref{eq:olivierkleinoop} for the ip (blue) and oop (red) magnetic field geometry. (b) Cooperativity $C = \nicefrac{g_\mathrm{eff}^2}{\eta_\mathrm{a}\kappa_\mathrm{s}}$ as a function of temperature for both geometries. The error bars originate from the covariance matrix of the least mean square fit.}
    \label{fig:g-C}
\end{figure}

We next discuss the magnon-phonon coupling rate in the context of magnon-phonon hybrids. The  magneto-elastic coupling constant $B_\mathrm{ip/ oop}$ is linked to the magneto-elastic coupling rate $g_\mathrm{eff}$ via \cite{An2020, An2023, Ye1988}

\begin{equation}
    \label{eq:olivierkleinip}
    g_\mathrm{eff, ip} (f) = \frac{B_\mathrm{ ip}}{2}\sqrt{\frac{g^2\mu_\mathrm{B}^2\mu_0H_\mathrm{res,ip}}{h^2 f^2 M_\mathrm{s}\Tilde{\rho_\mathrm{t}}d (L+d)}} \biggl [1-\mathrm{cos}\biggl(2\pi\frac{fd}{\Tilde{v_\mathrm{t}}} \biggr ) \biggr]
\end{equation}
and 
\begin{equation}
    \label{eq:olivierkleinoop}
    g_\mathrm{eff, oop} (f) = B_\mathrm{ oop}\sqrt{\frac{g\mu_\mathrm{B}}{h f M_\mathrm{s}\Tilde{\rho_\mathrm{t}}d (L+d)}} \biggl [1-\mathrm{cos}\biggl(2\pi\frac{fd}{\Tilde{v_\mathrm{t}}} \biggr ) \biggr].
\end{equation}
Here, $g$ is the Landé g-factor of YIG, $\mu_\mathrm{B}$ the Bohr magneton, $h$ the Planck constant, $M_\mathrm{s}$ the saturation magnetization of YIG and $\mu_0H_\mathrm{res, ip}$ the resonance field of the magnon Kittel mode at $f_0 = \SI{7}{GHz}$ for the in-plane configuration \footnote{Note, that the coupling rate is defined so that frequency difference between the normal modes is $2 g_\mathrm{eff}$, which is  in contrast to the definition used by An\,\textit{et.al} \cite{An2023}.}. Using the bbFMR information about the magnetic field dependence of the magnetic resonance frequency, we determine the effective magnetization $M_\mathrm{eff} = M_\mathrm{s} - 2K_\mathrm{u}/M_\mathrm{s}$, which contains the saturation magnetization and contributions of the uniaxial magnetic anisotropy. As discussed in Appendix\,\ref{Appendix: Ms and K1}, we extract from this data the effective magnetization $M_\mathrm{eff}$, which we use to calculate the effective coupling strengths in Eqs.\,\eqref{eq:olivierkleinip},\,\eqref{eq:olivierkleinoop}, instead of the saturation magnetization $M_\mathrm{s}$, since the uniaxial anisotropy is typically on the order of a few $\SI{}{mT}$ \cite{Serha2024} and therefore small compared to the saturation magnetization $M_\mathrm{s}$.

Figure\;\ref{fig:g-C} (a) shows the obtained magneto-elastic coupling rate $g_\mathrm{eff}$ for both geometries. Interestingly, the coupling rate increases by a factor of three when the temperature is increased from $\SI{70}{K}$ to $\SI{300}{K}$ for magnetic fields oriented along the oop direction. In contrast, $g_\mathrm{eff}$ for the ip magnetic field geometry shows an increase with decreasing temperature, with additional features between $\SI{5}{K}$ and $\SI{45}{K}$. Below $\SI{45}{K}$ the coupling rate decreases, reaches a minimum around $\SI{20}{K}$ and then strongly increases again down to $\SI{5}{K}$. This minimum in the coupling rate appears at about the same temperature, where the FMR linewidth of YIG often peaks due to slowly-relaxing rare-earth impurities\,\cite{Maier-Flaig2017b}. The coupling rate $g_\mathrm{eff}$, together with the damping rates $\kappa_\mathrm{s}$ and $\eta_\mathrm{a}$, allow us to quantify the temperature dependence of the cooperativity $C = \nicefrac{g_\mathrm{eff}^2}{\eta_\mathrm{a}\kappa_\mathrm{s}}$, for both field orientations (see Fig\;\ref{fig:g-C} (b)). For the oop magnetic field geometry, we find the maximum of $C$ around $\SI{160}{K}$. Below this temperature, $C$ decreases due to enhanced magnetic damping. For the ip magnetic field configuration, the maximum is found around $\SI{105}{K}$. The minimum of $C$ at approximately $\SI{25}{K}$ is attributed again to the increased magnetization damping of YIG. However, below $\SI{30}{K}$, $\kappa_\mathrm{s}$ is nearly constant, while $g_\mathrm{eff}$ still increases towards lower $T$ resulting in an increase of $C$ towards lower temperature.
\section{Conclusion}

We report on the temperature dependence of the magneto-elastic interaction in yttrium iron garnet/gadolinium gallium garnet high overtone bulk acoustic resonators.
The bulk acoustic wave resonator is realized as a bilayer comprising a yttrium iron garnet magnetic thin film and a single-crystalline gadolinium gallium garnet substrate. Making use of the model presented in Ref.\,\cite{Cherkasskii2025}, we fit our experimental data to extract the effective magneto-elastic coupling rate and the damping rates of the coupled subsystem. In particular, we determine the temperature dependence of these quantities between $\SI{5}{K}$ and $\SI{300}{K}$. We find that the magneto-elastic constant for the ip magnetic field configuration exceeds that for the oop magnetic field configuration below $\SI{150}{K}$, while the magnetic damping remains about constant in this temperature range. Neither the decrease for the oop magnetic field configuration nor the increase in the magneto-elastic coupling constant below $T = \SI{100}{K}$ for the ip magnetic field configuration is expected from the bulk literature values. Our experiments demonstrate that low-temperature experiments are conceptually feasible and enable high cooperativities and a reduced number of thermal noise quanta. The presented results can serve as guidance for the design of magnon-phonon-based signal transduction schemes as discussed in Refs.\,\cite{Engelhardt2022a, Graf2021,Rashedi2025}.

\section*{Acknowledgments}

We acknowledge financial support by the Deutsche Forschungsgemeinschaft (DFG, German Research Foundation) via the research unit CHiPS (541503763), Germany’s Excellence Strategy EXC-2111-390814868, and the Transregio ConQuMat (TRR 360 – 492547816). This research is part of the Munich Quantum Valley, which is supported by the Bavarian state government with funds from the Hightech Agenda Bayern Plus (Lighthouse Project NeQuS). 

\appendix

\section{Temperature dependence of the saturation magnetization and the uniaxial anisotropy}
\label{Appendix: Ms and K1}

\begin{figure}[tbh]
    \centering
    \includegraphics[width=0.49\textwidth]{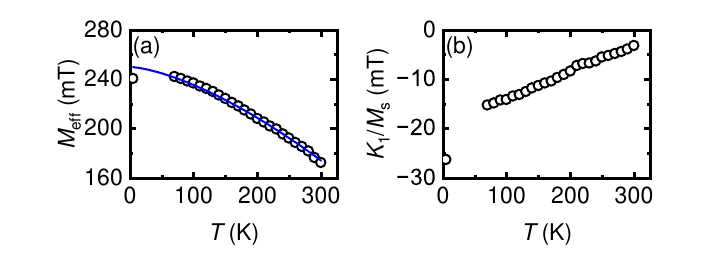}
    \caption{(a) Temperature dependence of the effective magnetization $M_\mathrm{eff}$. The blue continuous line represents a fit to the Bloch $T^{3/2}$ law. \cite{Bloch1930} (b) Temperature dependence of the cubic anisotropy field $K_1/M_\mathrm{s}$ in the YIG thin film.}
    \label{fig:Ms-K1}
\end{figure}

We fit the effective magnetization $M_\mathrm{eff}^\mathrm{ip,oop}$ for each geometric configuration (ip, oop), respectively. From these effective magnetizations and the resonance fields, we calculate the cubic anisotropy $K_1/M_\mathrm{s}$ using the following Kittel equations \cite{Kittel1948,Kohmoto2003,Lee2016,Dubs2020} for the ip and oop resonance frequency of a magnetic thin film as a function of the resonance magnetic field $\mu_0 H_\mathrm{res}^\mathrm{ip,oop}$

\begin{equation}
    \label{eq:oopKittel}
    f_\mathrm{res}^\mathrm{oop} = \frac{\gamma}{2\pi}\mu_0 \left(H_\mathrm{res}^{\mathrm{oop}} - M_\mathrm{s} - \frac{4K_1}{3M_\mathrm{s}}+\frac{2K_u}{M_\mathrm{s}}\right),
\end{equation}

and

\begin{equation}
    \label{eq:ipKittel}
    f_\mathrm{res}^\mathrm{ip} = \frac{\gamma}{2\pi}\mu_0 \sqrt{H_\mathrm{res}^{\mathrm{ip}}\left(H_\mathrm{res}^{\mathrm{ip}} + M_\mathrm{s} - \frac{K_1}{M_\mathrm{s}} - \frac{2K_u}{M_\mathrm{s}}\right)},
\end{equation}
where $M_\mathrm{eff}^\mathrm{oop} = M_\mathrm{s} + \frac{4K_1}{3M_\mathrm{s}} - \frac{2K_u}{M_\mathrm{s}}$ \\and 
$M_\mathrm{eff}^\mathrm{ip} = M_\mathrm{s} - \frac{K_1}{M_\mathrm{s}} - \frac{2K_u}{M_\mathrm{s}}$.  The ratio $K_1/M_\mathrm{s}$ displayed in Fig.\,\ref{fig:Ms-K1}\,(b) is determined via $\frac{3}{7}(M_\mathrm{eff}^\mathrm{oop}-M_\mathrm{eff}^\mathrm{ip})$.

\noindent Fig.\,\ref{fig:Ms-K1} shows the temperature dependence of an effective magnetization $M_\mathrm{eff} = \mu_0 M_\mathrm{s} - 2K_\mathrm{u}/M_\mathrm{s} = M_\mathrm{eff}^\mathrm{ip} + \frac{K_1}{M_\mathrm{s}}$, which is the saturation magnetization adjusted by the uniaxial anisotropy field as a function of temperature, which agrees well within $\SI{10}{\%}$ with previously reported values for the saturation magnetization \cite{Boventer2018}. This leads via propagation of uncertainty to a \SI{5}{\%} uncertainty for the ip and oop effective coupling strengths $g_\mathrm{eff}^\mathrm{ip, oop}$ in Fig.\,\ref{fig:g-C}. The blue continuous line represents a fit of the data to the Bloch $T^{3/2}$ law \cite{Bloch1930}

\begin{equation}
    \label{eq:Bloch}
    M_\mathrm{eff}(T) = M_\mathrm{eff}(0)\left(1-\left(\frac{T}{T_c}\right)^{3/2}\right),
\end{equation}
where $M_\mathrm{eff}(0) = \SI{250.1}{mT}$ and $T_c = \SI{665}{K}$ is the critical temperature approximating the $T^{3/2}$ dependence in considering the investigated temperature range. 
We use this fit to calculate the ip and oop effective coupling strengths $g_\mathrm{eff}^\mathrm{ip, oop}$ shown in Fig.\,\ref{fig:g-C}. 

\section{Used fitting function for the broadband ferromagnetic resonance}
\label{Appendix: fitting function}

We consider a bilayer system in which the non-magnetic layer occupies the region $d \le z \le L$, while the ferromagnetic layer extends over $0\le z \le d$. The magnetization dynamics in the ferromagnet is described by the Landau-Lifshitz--Gilbert (LLG) equation,
\begin{equation}
\partial_\mathrm{t}\boldsymbol{M}=-\gamma\mu_{0}\boldsymbol{M}\times\boldsymbol{H}^{\mathrm {eff}}+\frac{\alpha}{M_\mathrm{s}}\boldsymbol{M}\times\partial_\mathrm{t}\boldsymbol{M},
\end{equation}
where $\gamma$ is the gyromagnetic ratio, $\alpha$ is the Gilbert damping constant, and $M_{s}$ is the saturation magnetization. The effective magnetic field is defined via the functional derivative of the free-energy density $U$,
\begin{equation}
\boldsymbol{H}^{\mathrm{eff}}=-\dfrac{1}{\mu_{0}}\dfrac{\partial U}{\partial\boldsymbol{M}}.
\end{equation}
The total magnetic energy density in the ferromagnetic layer consists of several contributions. The Zeeman energy density is
\begin{equation}
U^{{\mathrm{Z}}}=-\mu_{0}\boldsymbol{H}\cdot\boldsymbol{M},
\end{equation}
and the demagnetization energy density is
\begin{equation}
U^{\mathrm{dm}}=\dfrac{1}{2}\mu_{0}\boldsymbol{M}\cdot\boldsymbol{N}\cdot\boldsymbol{M},
\end{equation}
where $\boldsymbol{N}$ denotes the demagnetization tensor. The magnetoelastic energy density is written as
\begin{equation}
U^{\mathrm{me}}=\frac{1}{M^{2}_\mathrm{s}}\sum_{ij}M_{i}M_{j}\left[B_{ij}\varepsilon_{ij}+K_{ij}\omega_{ij}\right],
\end{equation}
where $\varepsilon_{ij}$ is the strain tensor and $\omega_{ij}$ is the rotation tensor. The magnetoelastic coefficients are parameterized as $B_{ij}=\delta_{ij}B_{\parallel}+\left(1-\delta_{ij}\right)B_{\perp}$ with $B_{\parallel}$ and $B_{\perp}$ being the longitudinal and transverse magnetoelastic constants, respectively \footnote{In the main part the transverse magnetoelastic parameters are denoted as $B_\mathrm{oop}$ and $B_\mathrm{ip}$}. The magnetorotation coupling $K_{ij}$ originates from the rotation of the hard anisotropy axis.

The elastic free-energy density of the non-magnetic substrate is given by 
\begin{equation}
U^{{\mathrm{el}}}=\sum_{ijpq}\frac{\rho}{2}\left(\partial_\mathrm{t}u_{i}\right)^{2}+\frac{\varepsilon_{ij}c_{ijpq}\varepsilon_{pq}}{2},
\end{equation}
where $\rho$ is the mass density, $c_{ijpq}$ is the fourth-rank stiffness tensor, and the indices $i,j,p,q$ run over Cartesian coordinates. The elastic energy density of the ferromagnetic layer has the same structure, with the corresponding stiffness tensor and mass density denoted by tilted quantities.

To obtain the magnetic susceptibility, we expand the magnetic energy density up to quadratic order in the transverse components, using $M_\mathrm{z}=M_\mathrm{s}\left[1-\left(M^\mathrm{2}_{x}+M^{2}_\mathrm{y}\right)/2M^{2}_\mathrm{s}\right].$ We introduce circular variables $M_{\pm}=\frac{1}{\sqrt{2}}\left(M_\mathrm{x}\mp iM_\mathrm{y}\right),$$H_{\pm}=\frac{1}{\sqrt{2}}\left(H_\mathrm{x}\mp iH_\mathrm{y}\right)$. For an external magnetic field $\mu_{0}H_{0}$ applied perpendicular to the film plane (out-of-plane geometry), the magnetic susceptibility with respect to right-circularly polarized field reads
\begin{equation}
\chi^{{\mathrm{oop}}}=\frac{\omega_\mathrm{M_\mathrm{eff}^\mathrm{oop}}}{\omega_\mathrm{H}-\omega_\mathrm{M_\mathrm{eff}^{oop}}-i\kappa_\mathrm{s}-g^{{\mathrm{me}}}-\omega},
\end{equation}
For an external magnetic field applied in the film plane (in-plane geometry), the susceptibility becomes
\begin{widetext}
\begin{align}
\chi^{{\mathrm{ip}}} & =\frac{\omega_\mathrm{M_\mathrm{eff}^\mathrm{ip}}}{\left(\omega_\mathrm{H}-i\kappa_\mathrm{s}\right)\left(\omega_\mathrm{H}+\omega_\mathrm{M_\mathrm{eff}^\mathrm{ip}}-i\kappa_\mathrm{s}-g^{{\mathrm{me}}}\right)-\omega^{2}}\left(\omega_\mathrm{H}+\frac{\omega_\mathrm{M_{eff}^\mathrm{ip}}}{2}-i\kappa_\mathrm{s}-\frac{g^{{\mathrm{me}}}}{2}-\omega\right).
\end{align}
Here we introduced $\omega_\mathrm{M_\mathrm{eff}^\mathrm{oop,ip}}=\gamma\mu_{0}M_\mathrm{eff}^\mathrm{oop,ip},$ $\omega_\mathrm{H}=\gamma\mu_{0}H_\mathrm{ext},$ the magnetic damping rate $\kappa_\mathrm{s}$ and the magnetoelastic parameter 
\begin{align}
g^{{\rm me}} & =\frac{\gamma}{M_\mathrm{eff}^\mathrm{oop,ip}}\left(B_{\perp}+\frac{\mu_{0}\left(M_\mathrm{eff}^\mathrm{oop,ip}\right)^2}{2}\right)^{2}\frac{C_{4,4}k\sin\left(kL\right)\sin\left(\tilde{k}d\right)+4\tilde{C}_{4,4}\tilde{k}\cos\left(kL\right)\sin^{2}\left(\tilde{k}d/2\right)}{\tilde{C}_{4,4}\tilde{k}d\left[C_{4,4}k\sin\left(kL\right)\cos\left(\tilde{k}d\right)+\tilde{C}_{4,4}\tilde{k}\cos\left(kL\right)\sin\left(\tilde{k}d\right)\right]}.
\end{align}
\end{widetext}
The wave number of transverse elastic waves in the ferromagnetic layer is
\begin{equation}
\tilde{k}=\sqrt{\dfrac{\omega^{2}}{\tilde{v}^{2}_\mathrm{t}}+i\dfrac{2\tilde{\eta}_{\mathrm{a}}\omega}{\tilde{v}^{2}_\mathrm{t}}},
\end{equation}
while in the non-magnetic substrate it is 
\begin{equation}
k=\sqrt{\dfrac{\omega^{2}}{v^{2}_\mathrm{t}}+i\dfrac{2\eta_{\mathrm{a}}\omega}{v^{2}_\mathrm{t}}.}
\end{equation}
Here $c_{t}$ and $\tilde{c}_\mathrm{t}$ are the transverse sound velocities in the substrate and ferromagnet, respectively, and $\eta_{\mathrm{a}}$ and $\tilde{\eta}_{\mathrm{a}}$ denote the corresponding elastic damping coefficients. The stiffness components $C_{44}$ and $\tilde{C}_{44}$ are expressed in Voigt notation.


%

\end{document}